# Morphology effects on spin-dependent transport and recombination in polyfluorene thin films

Richards Miller[1], K. J. van Schooten[1], H. Malissa[1], G. Joshi[1], S. Jamali[1], J. M. Lupton[1,2] *, and C. Boehme[1] **

1. University of Utah, Department of Physics and Astronomy, Salt Lake City, Utah 84112-0830
2. Institut für Experimentelle und Angewandte Physik, Universität Regensburg, 93053 Regensburg, Germany

## Abstract

We have studied the role of spin-dependent processes on conductivity in polyfluorene (PFO) thin films by conducting continuous wave (c.w.) electrically detected magnetic resonance (EDMR) spectroscopy at temperatures between 10 K and 293 K using microwave frequencies between about 100 MHz and 20 GHz as well as pulsed EDMR at X-band (10 GHz). Variable frequency EDMR allows us to establish the role of spin-orbit coupling in spin-dependent processes whereas pulsed EDMR allows for the observation of coherent spin motion effects. We used PFO for this study in order to allow for the investigation of the effects of microscopic morphological ordering since this material can adopt two distinct intrachain morphologies: an amorphous (glassy) phase, in which monomer units are twisted with respect to each other, and an ordered (β) phase, where all monomers lie within one plane. In thin films of organic light-emitting diodes (OLEDs) the appearance of a particular phase can be controlled by deposition parameters and solvent vapor annealing, and is verified by electroluminescence spectroscopy. Under bipolar charge carrier injection conditions we conducted multi-frequency c.w. EDMR, electrically detected Rabi spin-beat experiments, Hahn-echo and inversion-recovery measurements. Coherent echo spectroscopy reveals electrically detected electron spin-echo envelope modulation (ESEEM) due to the precession of the carrier spins around the protons. Our results demonstrate that while conformational disorder can influence the observed EDMR signals, including the sign of the current changes on resonance as well as the magnitudes of local hyperfine fields and charge carrier spin-orbit interactions, it does not qualitatively affect the nature of spin-dependent transitions in this material. In both morphologies, we observe the presence of at least two different spin-dependent recombination processes. At 293 K and 10 K, polaron-pair recombination through weakly spin-spin coupled intermediate charge carrier pair states is dominant, while at low temperatures, additional signatures of spin-dependent charge transport through the interaction of polarons with triplet excitons are seen in the half-field resonance of a triplet spin-1 species. This additional contribution arises since triplet lifetimes are increased at lower temperatures.

* email: lupton@physics.utah.edu

** email: boehme@physics.utah.edu

I. INTRODUCTION

Spin-dependent electronic transitions of weakly coupled charge carrier pairs in organic semiconductors have been extensively investigated since they impart direct consequences for performance on optoelectronic devices made from these materials, such as organic light emitting diodes (OLED) [1]. These studies have led to significant insight into the role that various electron- and nuclear-spin related effects play in charge and spin transport within such devices [2–4]. Hyperfine interactions between charge carrier spins and their nuclear spin bath are particularly significant for correctly describing the microscopic origin of magneto-optoelectronic effects, such as organic magnetoresistance, and elementary parameters such as the spin-diffusion length. The latter is, for example, relevant for observing the inverse spin-Hall effect [5,6] in this material class. Although the local hyperfine environment has been found to significantly impact magnetoresistance behavior and imposes a limit on the sensitivity of magnetometry applications [7], little is currently known about how the microscopic magnetic environment is influenced by the local structure, i.e. the morphology of the polymer chain.

This study focuses on the qualitative nature and the dynamics of spin-dependent charge carrier processes in the polyfluorene derivative poly[9,9-dioctylfluorenyl-2,7-diyl] (PFO). PFO is an extraordinary conjugated polymer material for organic electronics in that the polymer chain can exist in two distinct conformations: a twisted conformation, in which the monomer units are free to rotate with respect to each other; and an ordered conformation, in which the side chains interlock so that all monomers lie within one plane [8]. The level of ordering can be followed by using, for example, x-ray scattering on ensemble bulk films, and has a dramatic impact on the optical properties such as photoluminescence and absorption. The conformational differentiation arises on the level of the individual chain which can be clearly illustrated by single-molecule spectroscopic techniques [8]. Thin films of this material can be prepared in different states of ordering, the amorphous glassy phase, the ordered β-phase and the mixed phases [9–11] by controlling deposition and processing conditions.

Here, PFO films are studied in OLEDs over temperatures between 10 K and 293 K under bipolar charge carrier (electron-hole) injection. In order to elucidate the influence of microscopic order on the fundamental spin interactions of electrostatically-bound charge carriers, we coherently probe spin-dependent recombination and dissociation kinetics in OLEDs through use of electrically-detected magnetic resonance (EDMR) spectroscopy [3]. With such marked changes to electronic characteristics between these two phases, a natural question arises: how does local structural order influence the spin degree of freedom for these charges? An effective method of probing spin-dependent electronic transitions in an operating OLED is to monitor the free charge-carrier density through the device current while performing electron spin resonant excitations on charge carriers. Since such spin manipulation of charge carriers involved in spin-dependent transitions leads to changes in overall device current [3], even coherent charge carrier spin propagation can be monitored through current measurements, allowing for electronic detection and quantification of electronic spin-spin coupling as well as hyperfine electron-nuclear spin interactions [4,12].

Polymer morphology has been explored extensively as a means of influencing the electronic performance of various organic electronic device schemes. This has been particularly true in the pursuit of efficient organic photovoltaic devices, where polymer composites of nominally electron accepting and donating materials are blended together, forming percolation networks, such as with $C_{60}$ and P3HT (poly[3-hexylthiophene-2,5-diyl]) [13–16]. One polymer system where structural order is known to lead to dramatic changes in electronic processes is PFO, which exhibits the above mentioned glassy phase and the rigid, molecular wire-like β-phase [17–20]. These two intramolecular conformation phases are sketched in the insets of Fig. 1(d, f). Although they share an identical chemical structure, the electronic properties of these two phases are distinctly different, leading to an order–of-magnitude conductivity enhancement [21] and a 15 nm luminescence red-shift from 425 to 440 nm in the β-phase [22]. The microscopic origins of such differences stem from the planar side chains of the wire-like π-conjugated backbone which forms an ordered ladder structure of the polymer in the β-phase [8,23]. With such dramatic changes in electronic properties between phases [19], PFO represents an ideal polymer for the study of morphological rather than mere chemical influences on spin properties.

Spin-dependent processes affecting the current in OLEDs arise primarily either from hyperfine or from spin-orbit interactions. Both effects could conceivably be influenced by molecular conformation. Intuitively, one may anticipate that the local strength of hyperfine fields increases in the β-phase as the local proton density is raised by ordering of the proton-rich side chains. However, this simple picture only applies if wavefunction localization is identical in both conformations – which is unlikely. In turn, twisting of the polymer chain, and the associated bending of the chain, may conceivably lead to stronger local spin-orbit coupling. But without quantitative spectroscopy it is impossible to make conclusive statements.

## II. EXPERIMENTAL PROCEDURE

Fig. 1(a) illustrates the OLED device structure used in this study. The structure is based on a glass template: indium tin oxide (ITO) (100 nm) is used as a transparent anode, poly(3,4-ethylenedioxythiophene) polystyrene sulfonate (PEDOT:PSS) is a hole injection material that is spin coated (at 3000 rpm) onto the cleaned ITO surface, and PFO in either morphology is the active layer that is spin coated in a nitrogen glovebox to form a thickness of approximately 75 nm. A thin (7 nm) calcium layer is then thermally evaporated to inject electrons, and 100 nm of aluminum is used to contact and encapsulate the device. The device is further encapsulated using either an epoxy or a so-called spin-on-glass (Futurrex IC1-200), for measurements at cryogenic temperatures, to ensure minimal atmospheric contact to the device during transport from the glovebox to the vacuum of a cryostat that is part of the spin resonance spectrometer.

Since PFO is a comparatively high mobility polymer, with mobilities ranging from $2 \times 10^{-9}$ to $3 \times 10^{-8}$ m²/Vs at 293 K depending on the morphological phase of the polymer [21], device performance can be negatively impacted by Joule heating for large active area devices at high current densities, causing a drop in the steady-state current for a fixed bias voltage. Therefore, in order to limit Joule heating, maintain device integrity, and increase layer homogeneity, a lithographically-structured SiN layer was used to define small circular openings to the underlying ITO with diameters of either 500 or 200 μm. This design reduces Joule heating by improved heat-sinking and therefore allows high-power device operation. Fig. 1(b, c) shows a scanning electron micrograph of the unprocessed active area and a photograph of light emission from the structured OLED, respectively. For all devices, the current-voltage characteristic was measured to ensure proper diode operation.

The degree of order within the PFO active layer can be controlled through layer deposition parameters and subsequent film treatment. A glassy-phase layer is formed by spin-coating a 5 g/L solution of the polymer in toluene directly onto the template for 1 minute with no further post-processing of the layer. PFO in the ordered β-phase is formed by briefly immersing a spin-coated layer of a slightly more concentrated solution (7 g/L in toluene) into a 1:1 orthogonal solvent mixture of tetrahydrofuran and methanol for 2 minutes, followed by thermal annealing at 100°C for 5 minutes, as outlined in Ref. [22].

To ensure absolute control over the morphological phase of the PFO OLEDs, the electroluminescence (EL) spectra were recorded for all devices. The EL spectrum can be used to infer long-range translational order within the polymer since the glassy phase exhibits emission for the 0-0 transition near wavelengths of 425 nm, while the β-phase has its maximal peak centered at 440 nm. This difference in 0-0 transition is a reliable indicator of the morphological phase of the layer [24]. Fig. 1(d-f) shows typical EL spectra for the glassy, mixed, and β-phases, respectively, along with illustrations of the polymer chains for each morphology.

In order to study the microscopic interplay between structural and magnetic order, we use pulsed-microwave EDMR (pEDMR) to coherently excite and observe spin transitions in the two phases of PFO. Experimentally, this is carried out by constructing OLEDs that have thin-film electrodes capable of being placed within the X-band (~9.6 GHz), low quality-factor Flexline MD5 microwave resonator of a Bruker ElexSys E580 spectrometer. Temperature control is achieved with a liquid helium cryostat (Oxford Instruments CF935O). When the microwave radiation is pulsed at 400 ns pulse duration and its frequency is kept fixed while the external magnetic field is swept across a range covering the resonance condition, a population transfer between spin eigenstates is induced, resulting in changes to the steady-state device current. The transient current changes were recorded using a Stanford Research Systems SR570 current amplifier [25].

All measurements were performed on devices with 200 μm diameter active areas under forward bias conditions so that a 20 μA forward current through the OLED was established. Exceptions were made for the experiments in Figs. 10-11, which had a steady state current of 50 μA to increase signal but also a larger active sample area in order to operate the material under similar conditions. We measured the change in current due to the magnetic resonant excitation by subtracting the steady-state current from the measured device current. A high-pass filter on the SR570 current amplifier with a cut-off frequency of 30 Hz was used to filter out low frequency contributions.

III. Results

A. Pulsed EDMR of PFO OLEDs

We begin by discussing the generic magnetic resonance signals observed in PFO OLEDs. Fig. 2 shows transient pEDMR signals for the two PFO phases, as well as for a device with an admixture of both phases, at both room temperature (293 K) and 10 K. Each panel depicts the change in current of a device from its steady-state current, encoded in the color scale, as a function of time after a 400 ns excitation pulse took place at time $t = 0$, plotted along the horizontal axis, and as a function of the applied magnetic field $B_0$, plotted along the vertical axis. The graphs to the right of each panel depict data subsets of the color plots that represent current changes as a function applied magnetic field recorded at times where maximal changes in current occurred, as indicted by the red arrows on the horizontal axis. Results from the β-phase sample [Fig. 2(a, b)] are shown above those from the mixed phase sample [Fig. 2(c, d)] and the glassy phase sample [Fig. 2(e, f)]; data measured at 10 K are shown to the right of the room temperature measurements. Note the different color scales exist for the different data sets. The current changes representing the magnitude of the detected spin-dependent currents increase 2 to 4 times when the devices are cooled to 10 K. Closer inspection of the transients reveals that each device shows a rather large change in current at shorter times, while a subsequent slower signal of opposite sign follows. The case in Fig. 2(d) is most instructive. Here, an initial enhancement in current is followed by a long-lived quenching. In the vicinity of the zero-crossing of the current change it can be seen that the quenching of the current envelopes the end of the initial enhancement peak around 40 μs after the microwave pulse: quenching and enhancement appear to occur simultaneously, implying that more than one spin-dependent mechanism must be active here.

Earlier studies on similar organic compounds point to the polaron pair (PP) model as the dominant origin of these transient magnetic resonance signals [26–31]. Typically, however, these earlier studies always showed a transient enhancement in current followed by a long-lived reduction (quenching). It is unusual that one and the same material can either show initial enhancement [e.g. Fig. 2(e)] or quenching [e.g. Fig. 2(a)]. Mixed-phase devices are particularly interesting since they show quenching at 293 K and enhancement at 10 K. This variation in current change can be explained by a change in transition rates, or the involvement of additional spin-dependent processes beyond the electron-hole PP process. One way to determine if there is more than one spin dependent channel (i.e. in addition to the PP process) is to fit the slice of the resonance spectrum at maximum amplitude (shown to the right of the colored panels in Fig. 2) with double Gaussian lines of equal area. Each charge in the pair should contribute to the resonance signal equally if the signal is due to a PP process. Each resonant spin (electron and hole) experiences inhomogeneous hyperfine broadening of slightly different magnitude, explaining the appearance of two Gaussians.

Figure 3 shows EDMR resonance spectra by plotting the change in current along the vertical axis as a function of magnetic field along the horizontal axis. The data are the same as shown in the insets of Fig. 2. Here, each plot shows the measured data (open squares) fitted with equal-area Gaussian curves (blue lines), and the combined fits of the two Gaussians (red lines). The residuals of each fit are shown above each panel. The data from the glassy phase device at 293 K (b) produces the only fit result that does not show a distinct structure in the residual given its signal-to-noise ratio. Structures in the residuals above the noise level imply that the two Gaussians do not provide a perfect fit to the spectra. Since these spectra offer only a snapshot of spectral broadening at one particular time after resonant spin excitation and for one particular magnetic field strength, we refrain from simply comparing spectral widths between the different phases for these measurements based on the data in Fig. 3. Instead, we will comprehensively discuss the magnetic field dependence of the resonance spectra below in Section C, which allows an extraction of the spectral line widths as a function of Zeeman splitting and offers a direct comparison between the two morphologies.

We first focus on the relation between chain morphology and the initial sign (quenching or enhancement) of the OLED current change following resonant excitation. The morphology can be quantified by considering the EL spectra. The inset in the upper left corner of Fig. 4 explains the procedure in relating the morphology to the initial sign of the transient current change. The inset illustrates the definition of two charges $A_1$ and $A_2$ as integrals of the current change in the time intervals between the resonant pulse and the sign change, and between the sign change and the relaxation of the current to the steady state (when the current change vanishes), respectively. Note that this inset is a sketch and not experimental data. Based on these charges, we introduce a normalization of $A_1$ by considering the ratio of $A_1$ to the sum of the magnitude of $A_1$ and $A_2$. This ratio represents the percentage of integrated charge that is due to initial current enhancement when the ratio is positive. When the ratio is negative, it represents initial current quenching. The main plot of Fig. 4 contains experimentally obtained values for this enhancement ratio (pEDMR experiments) for various PFO samples, prepared such that glassy, β-phase as well as the mixed phase emerged as a function of the morphology composition ratio detected in EL. The latter is an observable that is defined by the second inset in the lower right corner of Fig. 4, which shows a cartoon of an EL spectrum (see Fig. 1) which is fitted with two Gaussians whose integrated intensities $G$ and $\beta$ represent the EL intensities of the glassy ($G$) and β-phase, respectively. We now define the EL detected morphology composition ratio as $G/(|G|+|\beta|)$. The black line connecting the data points in Fig. 4 is a guide to the eye. Figure 4 shows that even with a significant glassy component in the EL spectra, the pEDMR signal sign can be dominated by current quenching characteristic of the β-phase. This observation is consistent with the observation made in Fig. 3 that β-phase EDMR signals are significantly stronger compared to those of glassy phase samples.

B. Half-field EDMR signals

Since the double-Gaussian fits shown in Fig. 3 do not lead to entirely vanishing residuals, we conclude that the PP process described by the double Gaussian function is not the only mechanism contributing to the signals observed in PFO. In order to identify these additional spin-dependent conductivity mechanisms, we carried out EDMR spectroscopy in the half magnetic-field domain in order to investigate the potential occurrence of a triplet-exciton polaron (TEP) process that has been observed before in other polymer films [32,33]. Triplet excitons can be quite short lived at room temperature, so their influence should be more pronounced at low temperatures [34]. The half-field measurements are conducted in the same manner as the measurements depicted in Fig. 2, but the magnetic field is set to slightly more than half the magnetic field where direct transitions between the sublevels of the triplets can become allowed for the given excitation pulse frequency.

While no half-field signal is observed in related conjugated polymers such as MEH-PPV at 293 K [34], PFO is known to have a potentially high triplet exciton density in the β-phase, owing to the longer triplet lifetime than for MEH-PPV, which could make room-temperature detection possible [35,36]. Fig. 5 shows the change in current as a function of magnetic field for β-phase (red triangles) and glassy phase (blue circles) OLEDs for both 293 K and 10 K. A half field resonance cannot be resolved for either phase at 293 K but is clearly visible at 10 K. Panels (c, d) show measurements at 10 K, where each resonance slice gives an average of multiple measurements to improve the signal to noise (4 averages were made for the β-phase and 6 for the glassy phase). The 293 K measurements were also averaged (22 averages for the β-phase, 9 averages for the glassy phase).

The half-field resonances can be fitted by a standard procedure using the EasySpin MATLAB toolbox in order to determine the zero-field splitting parameters $D$ and $E$, the dipolar and exchange coupling of the spin pair of the triplet exciton [37,38]. The errors for $D$ and $E$ were calculated using a bootstrap analysis [39]. The bootstrap method is a mathematical technique that uses the residuals of a fit to create new data sets with realistic noise values that in turn can be fitted again in an iterative process to give statistical distributions of the fit parameters. The β-phase zero-field splitting parameters of the triplet exciton are determined to be $D = 3216 \pm 81$ MHz and $E = 347 \pm 85$ MHz, while the glassy phase gave $D = 3560 \pm 270$ MHz and $E = 246 \pm 34$ MHz. Even though the amplitudes of the half-field resonances differ between the two morphological phases, the zero-field splitting parameters appear to be very similar. The residuals of the fits (black lines) in Fig. 5(c, d) are shown above the resonance curves. For the β-phase, it appears that there is some structure in the residual close to the resonance. Such a residual could be interpreted to imply that a third spin-dependent process, besides the PP and the TEP mechanisms, is present in PFO. This conclusion can also be drawn by noting that there is no detectable half-field resonance at 293 K in β-phase PFO, yet a clear structure in the fit residual of the full-field resonance is seen in Fig. 3(a). Even though the TEP mechanism is not detectable by the available EDMR experiment at room temperature, an additional non-dominant spin-dependent mechanism must exist.

C. Multi-frequency continuous wave EDMR and spin-orbit coupling

The line shape of the PFO full-field resonances shown in Figs. 2 and 3 is wider than that of similar polymers [12,40–44]. Inhomogeneous broadening of a magnetic resonance line of an amorphous material can result from local hyperfine interactions or a distribution in Landé g-factors which can arise from spin-orbit interactions. In order to determine how much of the resonance width is due to hyperfine coupling, arising from the abundance of hydrogen atoms in PFO, and what contribution results from spin-orbit coupling, multiple resonances were measured using coplanar waveguide resonators operating at different frequencies [45]. This approach is chosen because the two contributions to resonance line widths have different magnetic field dependencies: hyperfine broadening occurs independently of the external magnetic field strength, whereas spin-orbit coupling is manifested by a distribution of Landé-factors and therefore gains more influence on the spectrum for higher static magnetic field strengths.

For the multi-frequency EDMR experiments, the coplanar waveguide (CPW) resonators were operated under magnetic-field modulated continuous wave (cw) excitation as opposed to the pEDMR measurements discussed above. Consequently, the change in current as a function of magnetic field, i.e. the magnetic resonance spectrum, is recorded as a derivative function. Many frequencies are available in the CPW due to the use of the higher harmonics of each resonator's fundamental frequencies. This allows one to measure EDMR at a number of different magnetic fields and develop an understanding of how the line shape of the resonance changes with magnetic field strength. Fig. 6(a) shows the change in current due to multiple resonances for both glassy (blue) and β-phase (red) as a function of magnetic field (bottom horizontal axis) and corresponding frequency (upper axis). Since the magnetic field scale is so broad, the individual resonances appear very narrow.

As described in detail in Ref. [45] for a different polymer material, the multiple resonance spectra obtained through such a procedure can be analyzed by fitting all spectra simultaneously using a global fit with two field-dependent line widths $\Delta B_1$ and $\Delta B_2$ given by $\Delta B_{1,2} = \sqrt{B_{Hyp_{1,2}}^2 + (\alpha_{1,2} B_0)^2}$. This relation is based on a two-Gaussian model, representing one Gaussian function for the electron spin resonance and one for the hole spin resonance, which together form an ensemble of PPs undergoing spin-dependent recombination. This approach allows us to deduce the exact distribution width of the random hyperfine fields of each carrier 1 and 2 ($B_{Hyp_{1,2}}$) as well as the spin-orbit controlled Landé-factor distribution widths ($\alpha_{1,2}$). Fit results to the resonances for both polymer phases are shown in Fig. 6(b) for the lowest (upper curve) and highest (lower curve) frequencies recorded. The curves are shifted along the abscissa by the magnitude of the magnetic field on resonance, $B - f/\gamma$, with $\gamma$ being the gyromagnetic ratio. The fits are clearly of acceptable quality even though they are made under neglect of the above discussed second spin-dependent process and, more importantly, even though these global fit models are applied to a frequency span of a factor of 20. The fit quality does appear to deteriorate at higher frequencies for both phases of the material. The results of the fits are summarized in Table I. Note that the black fitting curve is actually one and the same fit for each phase, since a global fit is carried out over all data and the only variable in the plotted function is the magnetic field strength $B$.

We applied a bootstrap error analysis to the results of the global fits for both material phases to arrive at a better understanding of the uncertainty in the hyperfine and SO terms as described in detail in Ref. [45]. Fig. 6(c) shows the 95% confidence interval for the resonance line width ($\Delta B$) for both the narrow and wide Gaussian lines of the resonances (blue, glassy phase; red, β-phase) as a function of magnetic field. These errors extracted from the bootstrap analysis are stated in Table I. The circles and triangles in the plot mark the magnetic field strengths of each resonance spectrum taken in panel (a). The broader the resonance line, the larger the error in extrapolating the line width. Clearly, all features do indeed broaden with increasing magnetic field, implying contributions to line broadening from spin-orbit coupling resulting in a distribution $\Delta g$.

D. Detection of coherent spin motion with pEDMR

*1.    Rabi oscillations*

OLEDs have shown remarkable signatures of spin coherence such as spin beating between precessing electron and hole spins [46] and time-resolved electron-nuclear spin precession [4], phenomena which both show a strong dependence on hydrogen isotope. PFO is a unique material to investigate spin coherence effects since the two phases are chemically identical but structurally distinct: is there an effect of polymer structure on spin coherence? Fig. 7 shows the transient current response following a microwave pulse as a function of pulse duration for glassy and β-phase OLED devices. If the carrier spins retain their coherence, then spin-Rabi flopping becomes apparent in the device current [29]. Figs. 7(a, d, g, j) show the transient current response following a microwave pulse of varying duration. In order to improve the visibility of coherent Rabi oscillations as a function of microwave pulse length, the background was subtracted with a second-order polynomial function. This procedure is described in detail in Ref. [12]. The change in current is shown on a color scale as in Fig. 2. The amplitude $B_1$ of the excitation microwave pulse strength for the measurements in Fig. 7 is approximately 560 µT.

Panels (b, e, h, k) show slices along the respective white dashed lines to better portray the oscillation in device current as a function of excitation pulse length. The first few nanoseconds of each slice are omitted to better fit the data into the given scale for the displayed range of pulse lengths. As expected, all four data sets – for the two phases at the two temperatures – show coherent oscillations in the current. However, one can clearly see that the oscillations at 293 K decay more rapidly than those at 10 K.

The Rabi oscillations can be further analyzed by considering the frequency components making up the oscillation. Fig. 7(c, f, i, l) shows the Fourier transform of the time domain data for the time slices marked in white. To prevent distortions of the Fourier spectra by the baseline subtraction mentioned above, all transforms were carried on the uncorrected data without baseline subtraction. All Fourier spectra show a dominant fundamental at the Rabi frequency $\gamma B_1$, corresponding to the oscillation of one spin-½ carrier species. However, a second harmonic component is also seen at a frequency of $2\gamma B_1$, which arises due to simultaneous coherent precession – spin beating – of both the electron and hole spin. The detection of this spin-beating component is proof that the dominant spin-dependent transition for the observed EDMR signals is governed by weakly coupled pairs with spin s=1/2. This beating is consistent with the PP recombination mechanism which has previously been observed in MEH-PPV. The second-harmonic peaks in panels (c, i) are less pronounced than those in panels (f, l) because of the faster decay of the Rabi oscillations at room temperature. The beating component is, again, clearly visible in the 10 K data, and more so in the glassy phase than in the β-phase.

## 2.  Spin relaxation times

The Rabi oscillations on their own only demonstrate that coherent spin precession contributes to the device current under magnetic resonance excitation, but do not allow us to extract spin relaxation times. We use electrically detected Hahn-spin echo experiments and inversion recovery measurements to determine spin relaxation and dephasing times. We determine the necessary duration of the echo-driving π-pulse, which rotates the spins by 180° from their thermal equilibrium orientation along the direction of the external field $B_0$ to $-B_0$, from the duration of the Rabi oscillation at a given microwave power. Details of the echo experiments on OLEDs are given elsewhere [4,47]. Note that for the electrical detection of spin echoes, for which spin permutation symmetry rather than spin polarization is observed, it is necessary to modify the Hahn-echo pulse sequence well known for inductively detected magnetic resonance spectroscopy by adding an additional π/2-pulse. This pulse projects the charge carrier spins onto their eigenstates along the $\pm B_0$ axis. This procedure is explained in Ref. [4] and the *Supporting Information* thereof. Fig. 8 shows examples of Hahn echoes measured on PFO devices. The actual pulse sequence is illustrated above the figure, and the echo shape is recorded by varying the timing of the projection pulse ($\tau'$).

Figure 8 shows representative current-detected Hahn echoes for both PFO phases at room temperature and at 10 K (red shows the β-phase and blue the glassy phase). All data sets can be fitted with a simple Gaussian function which serves as a guide to the eye. The echoes measured at 10 K show a greater change in overall charge (i.e. time-integrated current) than those measured at room temperature. Figure 9 shows the decay of the echo envelope, i.e. the Hahn echo signal as a function of delay time $2\tau$ (with $\tau' = \tau$) at 293 K (a) and 10 K (b) for both β-phase (red triangles) and glassy phase (blue circles), allowing to determine the transverse spin-relaxation times $T_2$. A mixed-phase device (black pentagons) was also measured at 293 K to explore whether a distinct change in $T_2$ arises from a blend of phases. Since the glassy-phase devices are significantly more unstable than the β-phase samples, rapid measurements are necessary in order to maintain device integrity throughout a measurement. As a consequence, fewer data points were recorded for glassy phase devices. The signal-to-noise ratio of the 10 K β-phase measurement was low, and we therefore used a bootstrap error analysis in order to reliably determine upper and lower bounds for the $T_2$ values. The black lines show fits of single exponential decays for each data set. The resulting values for $T_2$, given in Table II, are of the same order of magnitude as those found in OLEDs made of other organic semiconductor molecules [48]. Similarly, very little effect of morphology is seen on the decoherence times $T_2$.

Spin relaxation is characterized both by the spin coherence time $T_2$ and the spin-lattice relaxation time $T_1$. In order to determine $T_1$ we conducted inversion recovery experiments. The pulse sequence is depicted above Fig. 10. A π-pulse is applied before a Hahn echo sequence, and the mixing time $T$ is varied. $T_1$ is usually longer than $T_2$ [48]. The data in Fig. 10 are plotted as the total detected charge as a function of mixing time $T$. The β-phase data are shown in red and glassy in blue with respective fits in black. Both data sets were recorded at 293 K. No measurements were carried out at 10 K. Since the β-phase OLED had a larger active area than what was used for the other measurements (500 μm diameter rather than 200 μm), a higher steady-state current of 50 μA was used to reach approximately the same current density as in the other measurements (~600 A/m$^2$). The extracted values for $T_1$ are summarized in Table II. Again, little difference is seen between the two phases.

*3. Electron-Spin-Echo Envelope Modulation*

In order to investigate the nature of the hyperfine couplings, which are responsible for the random effective magnetic fields governing the line width at low excitation frequencies, electron-spin-echo envelope modulation (ESEEM) measurements were performed. In echo measurements, the exponential decay of the echo amplitude, the echo envelope, is modulated slightly due to precession of the local nuclear magnetic moments in the course of the echo decay. Such modulations are not always observed, but when they arise, they provide a direct fingerprint of the dominant isotope responsible for hyperfine coupling. Figure 11 shows stimulated echo ESEEM experiments, following the procedure outlined in Ref. [4] with an illustration of the stimulated three-pulse echo sequence given above the figure. In such an experiment, nuclear polarization is generated by a $\pi/2$-$\tau$-$\pi/2$ pulse sequence acting on the electronic spins. The system then evolves freely for a mixing time $T$, and a stimulated echo is generated by another $\pi/2$ pulse. The final $\pi/2$ readout pulse is required for electrical detection when the spin-dependent current is governed by spin permutation symmetry rather than spin polarization [4]. The resulting echo signal at 293 K is shown in panel (a) for the two phases, where a Gaussian fit is again used as a guide to the eye. The stimulated echo is recorded as a function of mixing time $T$ with $\tau^* = \tau$ and is shown in panel (b).

The β-phase device used in this experiment was operated at a current of 50 μA due to this particular sample being manufactured with a larger active area. The device had a 500 μm diameter opening in the SiN insulating layer, as opposed to the 200 μm used for the glassy sample. As expected, the larger pixel produces a larger signal. However, larger pixels also displayed more of a tendency to random current fluctuations. The OLEDs based on glassy PFO with identical device structure could not maintain a sufficiently stable current, hence the small area template was used in the glassy ESEEM measurements. The ESEEM decay shown in Fig. 11(b), measured as time-integrated current (charge), depicts the charge decay as a function of mixing time, *T*. Fine structure in the measured stimulated echo decays shown in panel (b) might appear to be noise, however, it is not noise but a well-defined harmonic contribution caused by the precession of nuclear magnetic moments. This effect is clearly revealed by the Fourier transform of the decay curve shown in panel (c), where a distinct peak is seen around the frequency of 14.5 MHz for the β-phase OLED. This frequency corresponds exactly to the matrix proton frequency at the X-band field used, and therefore provides a clear demonstration of the hydrogen nuclei interacting with the PPs which in turn are responsible for conductivity. While the glassy-phase device shows a similar decay to the β-phase device in Fig. 11(b), there is no signal discernable at the hydrogen frequency in the Fourier transform in panel (c). However, this result does not allow for any conclusions, as the signal-to-noise ratio for an electrically detected ESEEM signal is less than unity for the glassy phase device (blue), below that of the more stable β-phase device. Given the chemical makeup of PFO, the fact that hyperfine-broadening appears to be larger in the glassy phase than in the β-phase (cf. Fig. 6) along with the general similarity of the spin-dependent processes in both phases strongly suggest that the charge carrier spins of the glassy phase should also experience oscillations close to 14.5 MHz. They are simply not detectable with the given experiment. This limitation can, in principle, be overcome by averaging over more repetitions of the experiment. However, we estimate that the duration of an experiment capable of resolving the hydrogen signal in the glassy device would exceed the lifetime of our devices in this phase. We note that it is well known from single-molecule spectroscopy that the β-phase is much more photostable than the glassy phase [8], so it is not surprising that the same also holds for devices.

IV. DISCUSSION

The results shown in Fig. 2 demonstrate that the microscopic ordering of the π-conjugated polymer PFO does indeed play a role for the quantitative nature of spin-dependent transitions in this material. The most significant effect is the sign change in resonantly induced current changes between amorphous (glassy) and ordered (β) phases. The magnitude of the current-transient response also differs substantially. Both phases show an increase in signal strength at 10 K for most measurements with respect to room temperature. The quenching occurring simultaneously with an enhancement, seen in Fig. 2(d) around 40 μs after the excitation pulse, indicates that there must be more than one single spin dependent process, since a single process can only give rise to either quenching or enhancement, but not both at the same time [3]. This conclusion is reaffirmed by the structure in the residuals of the Gaussian fits to the spectra in Fig. 3 for β-phase devices at 293 K (panel a) and 10 K (panel c) along with the glassy phase devices at 10 K (d). These fits also show that at least at room temperature there is one dominating spin-dependent electronic transition while the other spin-dependent mechanisms is significantly weaker in magnitude as corroborated by the weak (or absent) residuals at 293 K in Fig. 3a,b. For the dominant spin-dependent signal at room temperature, all evidence found in this study points towards the PP mechanism, the spin-dependent recombination of electrons and holes. These carriers first form weakly spin-coupled pairs due to Coulomb attraction before they recombine into singlet or triplet excitations, dependent on the PP spin state.

The reason for the inversion in sign in the transient current signal between the two types of devices (glassy or β-phase) in Fig. 2 is presently unclear. It is most certainly not due to the presence of the aforementioned additional spin-dependent channel. A straightforward hypothesis is that it originates from a change in balance of spin-dependent dissociation and recombination rates which describe the resonant current transients [49]. In Fig. 1(e) an EL spectrum for a mixed phase device is shown. There is a strong emission peak near 425 nm, characteristic of the glassy phase of PFO, while the second peak coincides with both the 0-1 glassy transition peak and the 0-0 emission of the β-phase. β-phase polymer chains are energetically favorable for charge carriers as indicated by the red shifted emission of the β-phase. Consequently, in a mixed-phase film, charge carriers will preferentially take paths through the active layer via β-phase chromophores rather than glassy-phase chromophores. Pure glassy-phase films are hard to achieve in device fabrication as the polymer will tend towards arranging itself in the preferential low-energy assembly of the β-phase. While a device can have a majority of glassy-phase polymer chains, remaining β-phase chains throughout the active layer can play a substantial role in the resulting EDMR signal. The color scales in Fig. 2 show that the β-phase devices display a much larger change in current when compared to the glassy phase devices. Hence, a small percentage of β-phase polymer chains within a predominantly glassy-phase film can have a larger effect than the predominantly glassy-phase chains and burry the signal originating from the glassy phase. Fig. 2(c, d) shows a change in signal sign between 293 and 10 K measurements of the mixed phase film, as well as relatively weak changes in current per excitation pulse compared to pure β-phase devices. The spin-dependent properties of the mixed phase devices therefore indeed lie between the two extremes of glassy and β-phase.

At 10 K, a TEP process is seen in both phases of PFO due to the appearance of a half-field resonance, but the PP process remains the dominant spin-dependent mechanism for the polymer at both temperatures studied. The dominance of the PP process is demonstrated by the spectral analysis of the magnetic-field slice of the resonance and the appearance of Rabi spin-beating at a frequency of $2\gamma B_1$. There may potentially be another spin-dependent transport process present in PFO that was not discerned in this study. The zero-field splitting parameters of the triplet exciton are found to be very similar for the two phases, implying a similar degree of localization of the triplet. However, the TEP process appears to be more prominent in the β-phase, which likely relates to the longer triplet lifetime in the ordered material, in agreement with earlier studies of photoinduced absorption [17].

In both polymer conformations, the line width in the limit of low excitation frequencies is determined solely by hyperfine coupling between charge carrier spins and hydrogen nuclei. It is significantly larger than what has been observed in other organic semiconductor materials [12,26,27,42]. The line width increases with excitation frequency, which also points to a strong contribution of spin-orbit coupling. The glassy-phase devices exhibited larger hyperfine broadening but a slightly smaller spin-orbit term than the ordered β-phase. One possible explanation for the counterintuitive observation of stronger hyperfine coupling in the glassy phase devices (where the proton density is lower due to the disorder) is a stronger localization and broader conformational and energetic distribution of the possible charge carrier states. A related effect is known from amorphous Si:P, where the field splitting of the hyperfine lines is 24 mT as compared to 4.2 mT in crystalline Si:P [50]. The double Gaussian derivative line shape used to globally fit the c.w. EDMR resonances describes the measured spectra well at low excitation frequencies, but deviates progressively for higher frequencies. The individual fits for the glassy-phase resonances went from fit qualities of $R^2 = 0.966$ at 2.33 GHz to $R^2 = 0.712$ at 17.89 GHz. The β-phase at 1.15 GHz fitted with $R^2 = 0.982$ but only $R^2 = 0.852$ at 19.88 GHz. These deviations at higher frequencies could be partially due to the spin-dependent processes not accommodated by the PP model, i.e. possibly a TEP process. However, these mechanisms contribute equally to lower-frequency spectra where excellent fit agreements have been achieved. It is therefore more likely that the deteriorating fits at higher frequencies are caused by the inadequacy of the assumption that the *g*-factors are isotropic. As spin-orbit contributions become increasingly significant at spectra recorded at higher frequencies, the *g*-factors may increasingly require representation by the full *g*-tensors. Whether or not this hypothesis is correct, though, can only be resolved experimentally through EDMR experiments at frequencies much greater than 20 GHz.

Even with the imperfect two-Gaussian model for the global fit, the results of the bootstrap analysis (see Table I) show a significant increase of hyperfine field strength in the glassy phase compared to the β-phase. The distributions of the *g*-factors *Δg* are all very similar except for the order-of-magnitude greater value seen in the broad resonance peak for the β-phase. We therefore conclude that there is likely very little influence of chain shape on spin-orbit coupling. However, chain shape does affect hyperfine coupling: in the more disordered material, proton densities can increase locally, raising local hyperfine fields and inducing spectral broadening.

PFO devices are particularly instructive for understanding the relation between spin-coherence, as revealed by electrically detected spin-Rabi oscillations, and the underlying spin coherence times, which can be extracted using Hahn echoes. Although substantial differences in the fidelity of Rabi oscillations appear to exist, with the highest-quality oscillations apparent in the 10 K glassy-phase data, there is very little quantitative difference in the spin lifetimes. The glassy phase shows a charge-carrier spin coherence time of $T_2 = 237 \pm 37$ ns with a spin lattice relaxation time $T_1 = 5.6 \pm 0.9$ μs, both at 293 K. The β-phase coherence time was marginally longer at $T_2 = 295 \pm 10$ ns; the spin lattice relaxation time of the β-phase film was also slightly larger, $T_1 = 9.1 \pm 2.5$ μs. This similarity in values is consistent with the assumption that decoherence is driven by hyperfine interactions [47], since hyperfine coupling strengths are within the same orders of magnitude even though detectable differences in hyperfine-field strengths exist between the two phases.

The role of hyperfine coupling becomes particularly clear in the ESEEM experiments where we observe an echo envelope modulation signal corresponding to the hyperfine coupling to hydrogen nuclei. This spectral signature has been confirmed only in β-phase samples, yet the absence in the glassy-phase devices is probably related to the low signal-to-noise ratio and poorer device stability.

IV. SUMMARY AND OUTLOOK

This study has shown that OLEDs made out glassy and β-phase PFO display spin-dependent charge transport properties. This comparison probes only the influence of molecular morphology on spin-dependent processes without any variation to chemical structure. Surprisingly, the sign of the initial change for current transients is opposite between phases, implying that the balance between spin-dependent recombination and dissociation rates within a device structure is altered.

All coherence effects are similar for the two phases at both high and low temperatures. The zero-field splitting parameters are similar despite the TEP process being more prominent in the β-phase. The TEP process is present in both phases, but the PP process is the dominant origin of spin-dependent currents. The main differences that occur when the molecular order is altered is that the spin-resonance induced current change from the steady-state is greater for the ordered phase compared to the amorphous phase, while the hyperfine fields experienced by the charge carriers in the ordered phase are 63-76% as strong as those in the glassy phase. The spin-orbit related Landé-factor distributions are similar for the narrow line of the double-Gaussian resonance spectra, while for the broad resonance lines, the β-phase devices show about twice as broad distributions as the glassy phase. This observation suggest that besides the chemical structure, the conformation of an organic semiconductor material can, in principle, be used to tune hyperfine field strengths as well as spin-orbit coupling. Such a possible dependence of spin-orbit coupling on conformation in molecular semiconductors is extremely interesting but can only be resolved conclusively once $\Delta g$-broadening of the spectra substantially exceeds the hyperfine broadening. This will require resonance frequencies much higher than the 20 GHz used in the present work.

ACKNOWLEDGMENTS

This work was supported by the US Department of Energy, Office of Basic Energy Sciences, Division of Materials Sciences and Engineering under Award #DE-SC0000909.

V. REFERENCES

[1]  S. Geschwind, R. J. Collins, and A. L. Schawlow, Phys. Rev. Lett. **3**, 545 (1959).

[2]  J. Klein and R. Voltz, Phys. Rev. Lett. **36**, 1214 (1976).

[3]  C. Boehme and K. Lips, Phys. Rev. B **68**, 245105 (2003).

[4]  H. Malissa, M. Kavand, D. P. Waters, K. J. van Schooten, P. L. Burn, Z. V. Vardeny, B. Saam, J. M. Lupton, and C. Boehme, Science **345**, 1487 (2014).

[5]  G. E. W. Bauer, E. Saitoh, and B. J. van Wees, Nat. Mater. **11**, 391 (2012).

[6]  D. Sun, K. J. van Schooten, M. Kavand, H. Malissa, C. Zhang, M. Groesbeck, C. Boehme, and Z. Valy Vardeny, Nat. Mater. **20**, doi:10.1038/nmat4618 (2016).

[7]  T. D. Nguyen, G. Hukic-Markosian, F. Wang, L. Wojcik, X.-G. Li, E. Ehrenfreund, and Z. V. Vardeny, Nat. Mater. **9**, 345 (2010).

[8]  T. Adachi, J. Vogelsang, and J. M. Lupton, J. Phys. Chem. Lett. **5**, 2165 (2014).

[9]  A. L. T. Khan, P. Sreearunothai, L. M. Herz, M. J. Banach, and A. Köhler, Phys. Rev. B **69**, 085201 (2004).

[10] W. Chunwaschirasiri, B. Tanto, D. L. Huber, and M. J. Winokur, Phys. Rev. Lett. **94**, 107402 (2005).

[11] A. J. C. Kuehne, M. Kaiser, A. R. Mackintosh, B. H. Wallikewitz, D. Hertel, R. A. Pethrick, and K. Meerholz, Adv. Funct. Mater. **21**, 2564 (2011).

[12] K. J. van Schooten, D. L. Baird, M. E. Limes, J. M. Lupton, and C. Boehme, Nat. Commun. **6**, 6688 (2015).

[13] J. S. Kim, Y. Park, D. Y. Lee, J. H. Lee, J. H. Park, J. K. Kim, and K. Cho, Adv. Funct. Mater. **20**, 540 (2010).

[14] S.-S. Kim, J. Jo, C. Chun, J.-C. Hong, and D.-Y. Kim, J. Photochem. Photobiol. A Chem. **188**, 364 (2007).

[15] F. Yang, M. Shtein, and S. R. Forrest, Nat. Mater. **4**, 37 (2004).

[16] S. M. Lindner, S. Hüttner, A. Chiche, M. Thelakkat, and G. Krausch, Angew. Chem. **45**, 3364 (2006).

[17] A. J. Cadby, P. A. Lane, H. Mellor, S. J. Martin, M. Grell, C. Giebeler, D. D. C. Bradley, M. Wohlgenannt, C. An, and Z. V. Vardeny, Phys. Rev. B **62**, 15604 (2000).

[18] M. Ariu, D. G. Lidzey, M. Sims, A. J. Cadby, P. A. Lane, and D. D. C. Bradley, J. Phys. Condens. Matter **14**, 9975 (2002).

[19] K. Becker and J. M. Lupton, J. Am. Chem. Soc. **127**, 7306 (2005).

[20] E. Da Como, K. Becker, J. Feldmann, and J. M. Lupton, Nano Lett. **7**, 2993 (2007).

[21]   H. T. Nicolai, G. A. H. Wetzelaer, M. Kuik, A. J. Kronemeijer, B. de Boer, and P. W. M. Blom, Appl. Phys. Lett. **96**, 172107 (2010).

[22]   T.-H. Jen, K.-K. Wang, and S.-A. Chen, Polymer **53**, 5850 (2012).

[23]   M. Grell, D. D. C. Bradley, G. Ungar, J. Hill, and K. S. Whitehead, Macromolecules **32**, 5810 (1999).

[24]   B. Arredondo, B. Romero, a. Gutiérrez-Llorente, a. I. Martínez, a. L. Álvarez, X. Quintana, and J. M. Otón, Solid. State. Electron. **61**, 46 (2011).

[25]   M. Kavand, D. Baird, K. Van Schooten, H. Malissa, J. M. Lupton, and C. Boehme, arXiv:1606.05680 [cond-mat.mes-hall] (2016).

[26]   K. Murata, Y. Shimoi, S. Abe, S. Kuroda, T. Noguchi, and T. Ohnishi, Chem. Phys. **227**, 191 (1998).

[27]   G. B. Silva, L. F. Santos, R. M. Faria, and C. F. O. Graeff, Phys. B Condens. Matter **308-310**, 1078 (2001).

[28]   C. G. Yang, E. Ehrenfreund, F. Wang, T. Drori, and Z. V. Vardeny, Phys. Rev. B **78**, 205312 (2008).

[29]   D. R. McCamey, H. a Seipel, S.-Y. Paik, M. J. Walter, N. J. Borys, J. M. Lupton, and C. Boehme, Nat. Mater. **7**, 723 (2008).

[30]   L. S. Swanson, J. Shinar, A. R. Brown, D. D. C. Bradley, R. H. Friend, P. L. Burn, A. Kraft, and A. B. Holmes, Phys. Rev. B **46**, 15072 (1992).

[31]   V. A. Dediu, L. E. Hueso, I. Bergenti, and C. Taliani, Nat. Mater. **8**, 707 (2009).

[32]   T. L. Keevers, W. J. Baker, and D. R. McCamey, Phys. Rev. B **91**, 205206 (2015).

[33]   W. J. Baker, T. L. Keevers, C. Boehme, and D. R. McCamey, Phys. Rev. B **92**, 041201 (2015).

[34]   W. J. Baker, D. R. McCamey, K. J. van Schooten, J. M. Lupton, and C. Boehme, Phys. Rev. B **84**, 165205 (2011).

[35]   C. Rothe and A. P. Monkman, Phys. Rev. B **68**, 075208 (2003).

[36]   C. Rothe, S. King, F. Dias, and a. Monkman, Phys. Rev. B **70**, 195213 (2004).

[37]   S. Stoll, Spectral Simulations in Solid-State Electron Paramagnetic Resonance, ETH Zürich, 2003.

[38]   S. Stoll and A. Schweiger, J. Magn. Reson. **178**, 42 (2006).

[39]   B. Efron and T. Robert J, *An Introduction to the Bootstrap*, 1st ed. (Chapman & Hall, New York, 1993).

[40]   C. Boehme, D. R. McCamey, K. J. van Schooten, W. J. Baker, S.-Y. Lee, S.-Y. Paik, and J. M. Lupton, Phys. Status Solidi **246**, 2750 (2009).

[41]   B. Z. Tedlla, F. Zhu, M. Cox, J. Drijkoningen, J. Manca, B. Koopmans, and E. Goovaerts, Adv. Energy Mater. **5**, 1401109 (2015).

[42]   G. Li, C. H. Kim, P. A. Lane, and J. Shinar, Phys. Rev. B **69**, 165311 (2004).

[43] J. Niklas, K. L. Mardis, B. P. Banks, G. M. Grooms, A. Sperlich, V. Dyakonov, S. Beaupré, M. Leclerc, T. Xu, L. Yu, and O. G. Poluektov, **15**, 9562 (2013).

[44] K. Kanemoto, H. Matsuoka, Y. Ueda, K. Takemoto, K. Kimura, and H. Hashimoto, Phys. Rev. B **86**, 125201 (2012).

[45] G. Joshi, R. Miller, L. Ogden, M. Kavand, S. Jamali, K. Ambal, S. Venkatesh, D. Schurig, H. Malissa, J. M. Lupton, and C. Boehme, arXiv: 1603.03807 (2016).

[46] D. R. McCamey, K. J. van Schooten, W. J. Baker, S.-Y. Lee, S.-Y. Paik, J. M. Lupton, and C. Boehme, Phys. Rev. Lett. **104**, 017601 (2010).

[47] W. J. Baker, T. L. Keevers, J. M. Lupton, D. R. McCamey, and C. Boehme, Phys. Rev. Lett. **108**, 267601 (2012).

[48] S.-Y. Paik, S.-Y. Lee, W. J. Baker, D. R. McCamey, and C. Boehme, Phys. Rev. B **81**, 075214 (2010).

[49] D. R. McCamey, S.-Y. Lee, S.-Y. Paik, J. M. Lupton, and C. Boehme, Phys. Rev. B **82**, 125206 (2010).

[50] M. Stutzmann and R. Street, Phys. Rev. Lett. **54**, 1836 (1985).

# FIGURE 1

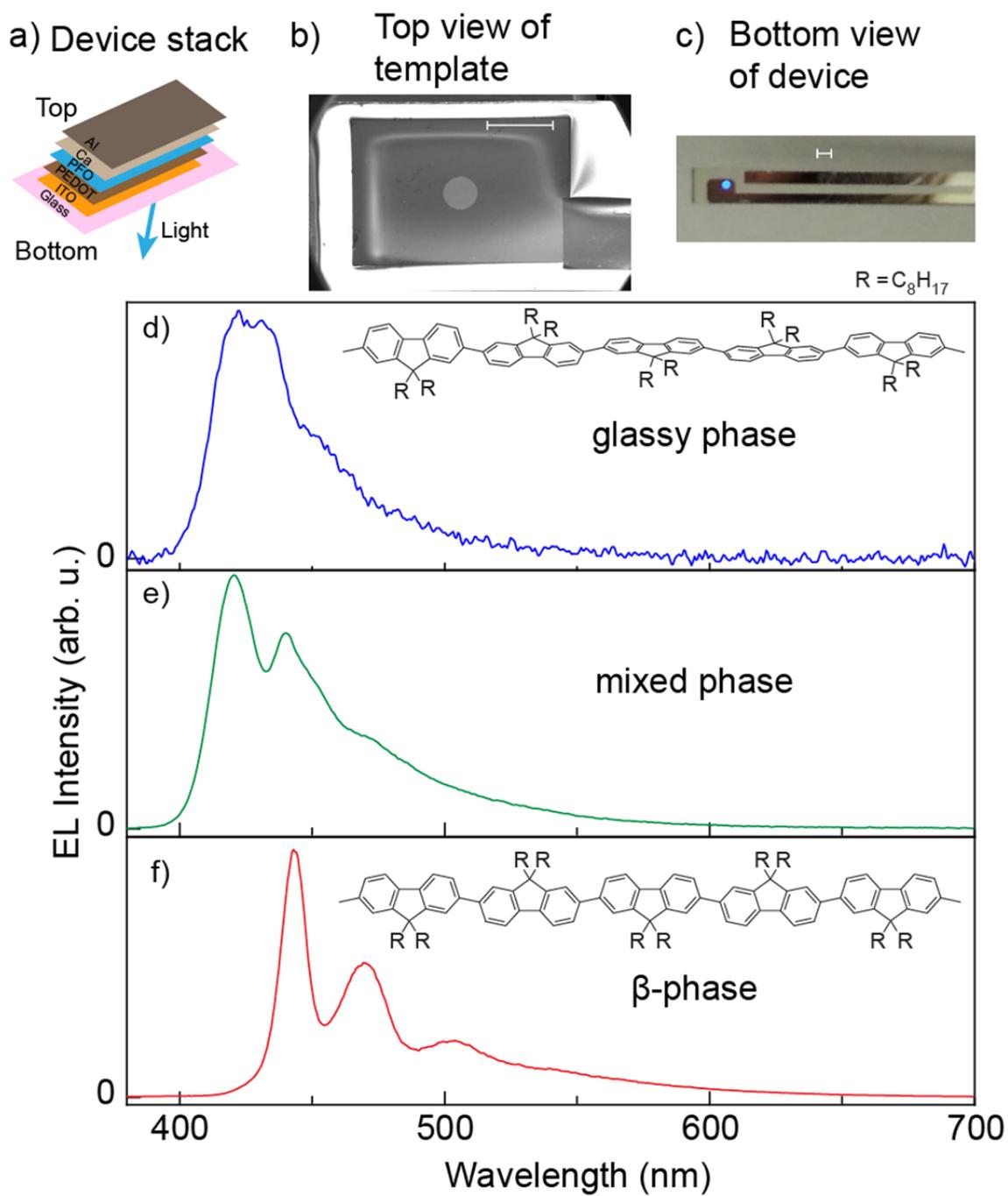

FIGURE 1: (a) Illustration of the vertical device stack used in all measurements of this study where the active layer of PFO (blue) is sandwiched between a Ca layer (light grey) and a PEDOT:PSS layer (brown) for electron and hole injection, respectively. (b) SEM image of the active area of a standard device covered with SiN to insulate the ITO except for a small circular opening in the center, defining the active area. The small active device area atop the large substrate allowed for sufficient heat sinking of the power dissipated by the device under operation. (c) Photograph of a device under operating conditions with the blue PFO EL visible. The scale bars in (b) and (c) both represent 1 mm. The EL spectra for glassy, mixed and β-phase devices are plotted in (d), (e), and (f), respectively. Sketches of the polymer chain conformations for the two main phases are shown in their respective panels in order to illustrate the different states of molecular ordering.

# FIGURE 2

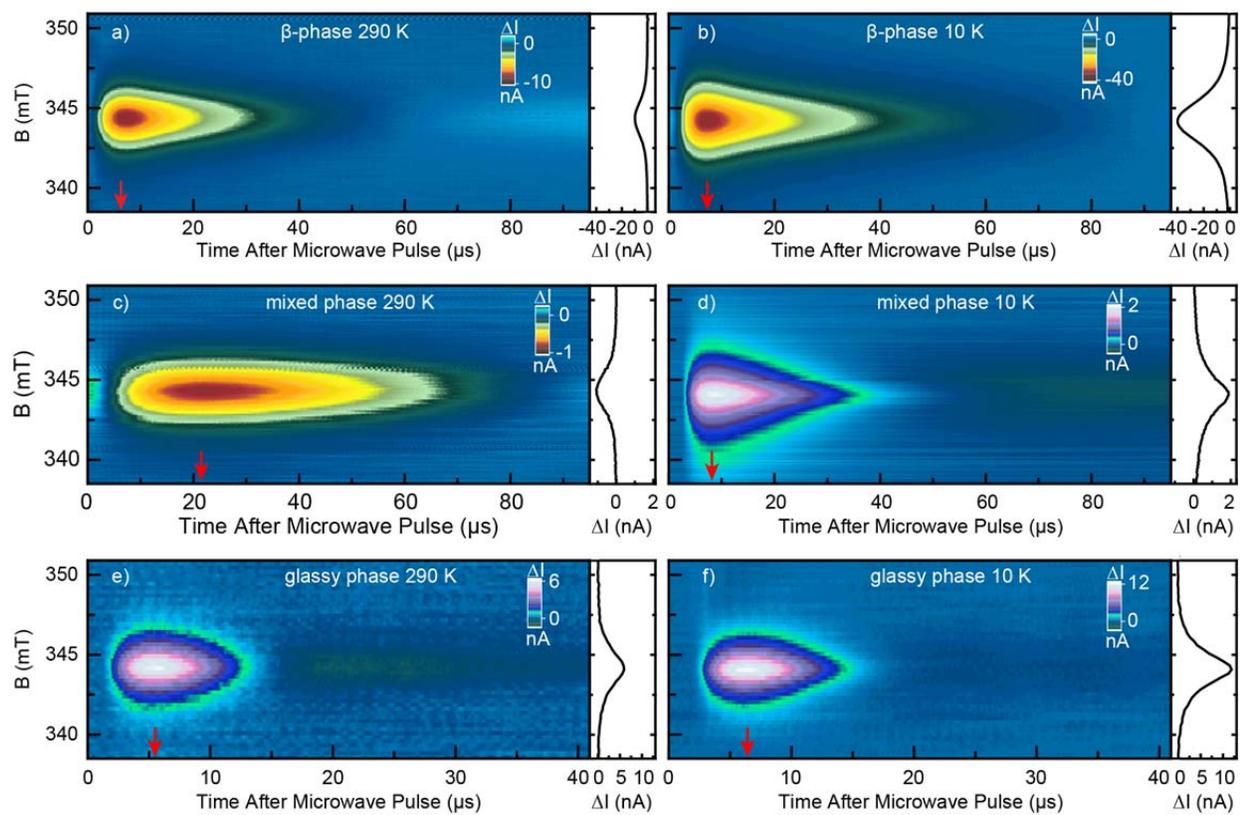

FIGURE 2: Plots of transient changes to a steady-state device current of $I_0=20$ µA after a 400 ns microwave excitation pulse was applied to the samples, as a function of the applied magnetic field which is represented by the vertical axis. The current transients were measured at 293 K (left) and 10 K (right), on devices containing PFO in the β-phase (a, b), a mixed phase containing both β-phase and glassy components (c, d) and the glassy phase (e, f). The insets of the panels display plots of the changes in device current as a function of the applied magnetic field for specific times after the microwave pulse indicated by the red arrows. Qualitatively, a sign reversal of the current change is seen between glassy and β-phases while for the mixed phase device a sign change occurs between high and low temperature. The data also show that magnetic resonance induced current changes in PFO are more than a factor of two larger in β-phase PFO compared to glassy PFO.

# FIGURE 3

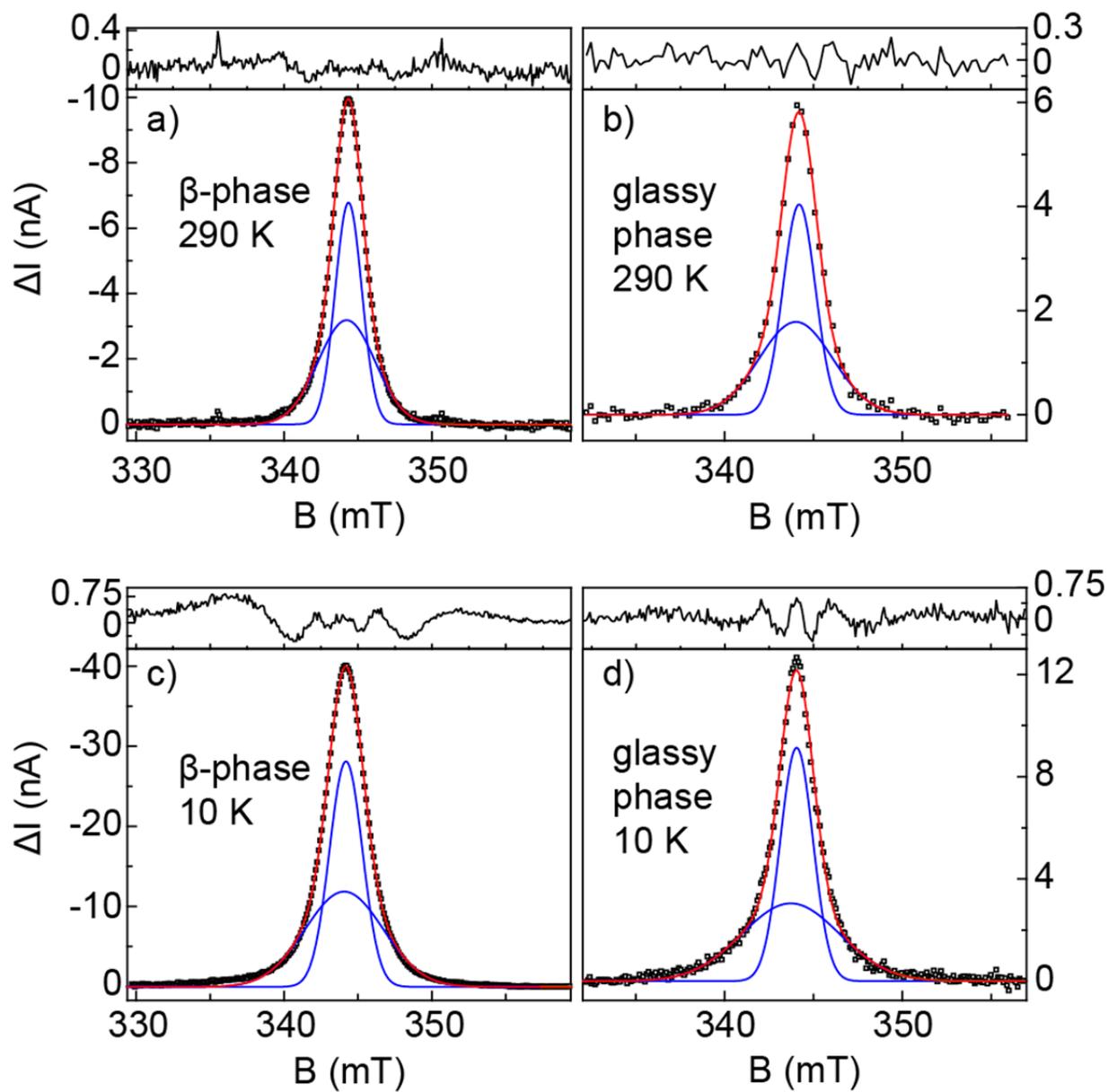

FIGURE 3: Plots of the measured maximal current change (black squares) after the resonant pulse excitation as a function of applied static field $B_0$ obtained from the data shown in Fig. 2 for β- and glassy phases at temperatures of 293 K (a, b) and 10 K (c, d). The blue lines are fit results with double Gaussian functions, representing electron and hole, in which both functions have the same area as required for a pair process. The quality of the fit results is recognized by the fit residuals which are plotted in the insets of the panels. For all residual data sets displayed, weak but significant structure is discernible, indicating that additional spin-dependent processes not described by a double Gaussian line make minor contributions to the overall EDMR response of the devices.

# FIGURE 4

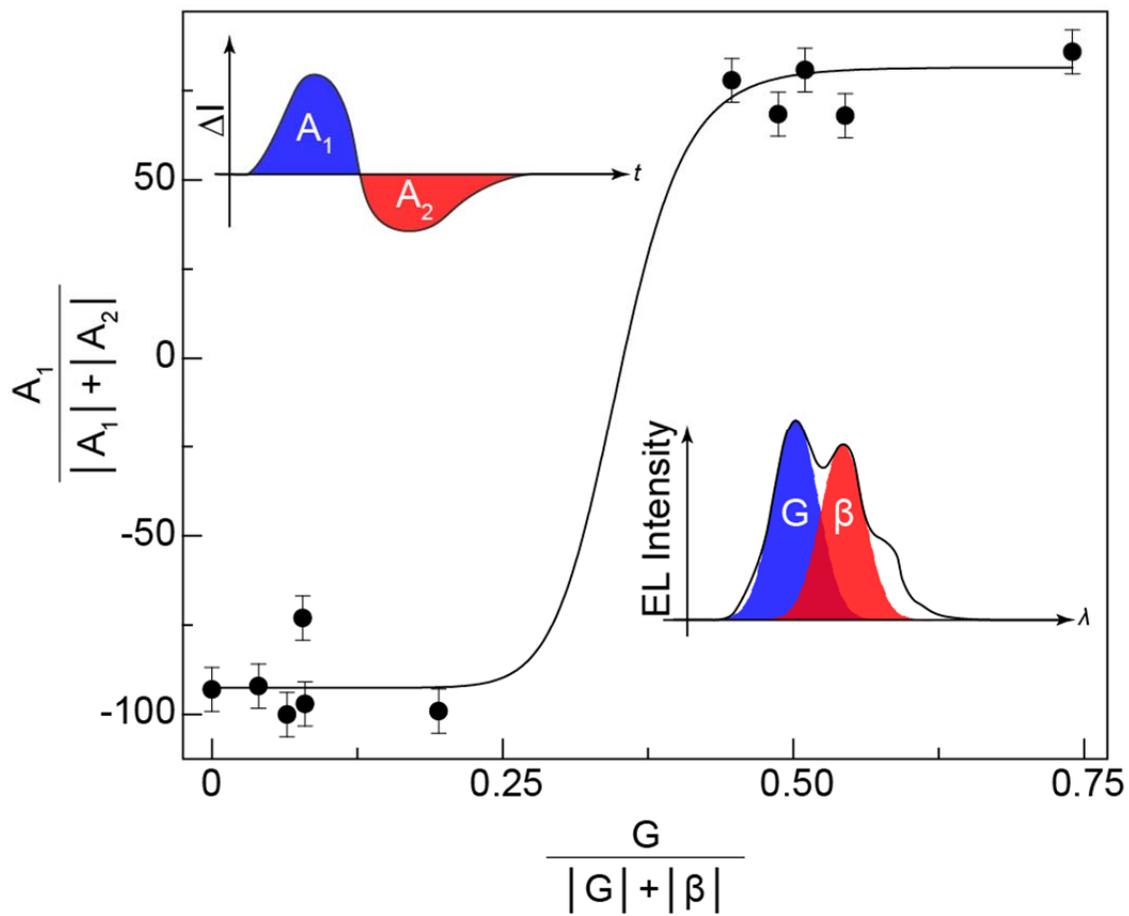

FIGURE 4: Correlation between EL spectrum and the sign of the initial steady-state current change following magnetic resonant excitation. The transient of current change is separated into two parts by its zero-crossing and integrated to yield areas $A_1$ and $A_2$ as sketched in the top left inset (this example transient is not experimental data). The ratio of $A_1$ to the sum of the moduli $|A_1|+|A_2|$ gives a measure of the sign of the initial part of the transient, either current enhancement or quenching. This ratio is related to the fraction of glassy-phase EL which is derived by deconvoluting the EL spectrum into glassy emission and β-phase emission. The fraction of the glassy phase is defined as the ratio between the glassy spectral EL peak intensity $G$ divided by the total EL intensity $|G|+|\beta|$, as illustrated in the bottom right inset. The black line is a guide to the eye. The data reveal that the sign of the steady-state current change of PFO OLEDs is governed by the fraction of glassy to β-phase.

# FIGURE 5

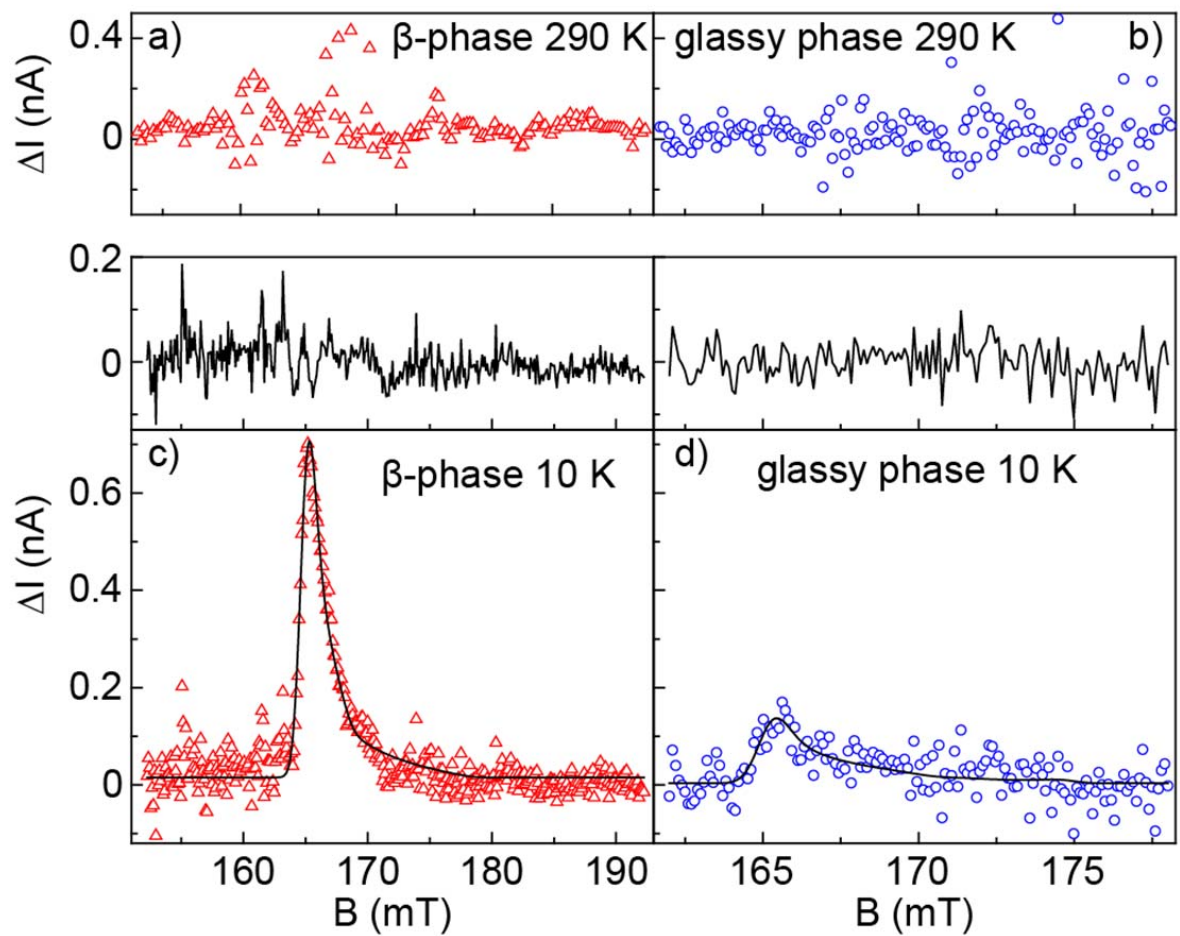

FIGURE 5: Plots of half-field resonance spectra for $\beta$-phase (red triangles) and glassy-phase (blue circles) devices at 293 K (a, b) and 10 K (c, d). Each panel shows a change in steady-state current as a function of magnetic field. The fits in (c, d) were calculated using the EasySpin MATLAB toolbox in order to determine the zero-field splitting parameters $D$ and $E$. A bootstrap error analysis was used to establish the error in the fit parameters. No discernable signal was found for either phase at 293 K.

# FIGURE 6

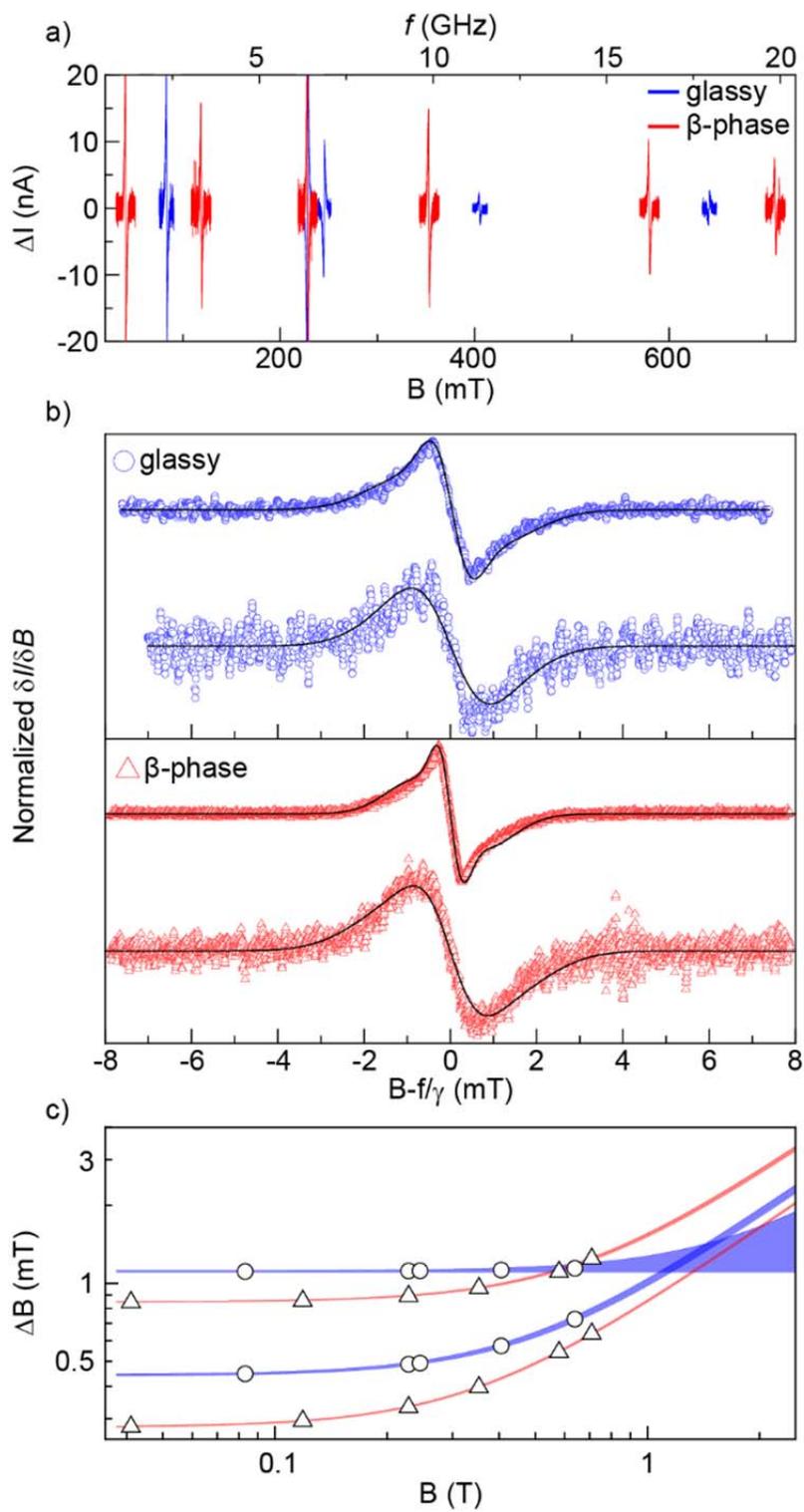

FIGURE 6: Multi-frequency continuous wave (c. w.) EDMR spectra obtained by using coplanar waveguide (CPW) resonators. (a) C. w. spectra of glassy and β-phase OLEDs, shown in blue and red, respectively. The change in current is shown as a function of both static magnetic field (bottom) and the corresponding magnetic resonance frequency for an uncoupled electron with the vacuum Landé-factor (top axis). Note that c. w. spectra have a differential lineshape compared to the pulsed spectra in Fig. 3. (b) Plot of normalized resonance spectra for different magnetic field scales as a function of the offset relative to the observed resonance center, obtained from devices with glassy (blue circles) and β-phase (red triangles) PFO. For each phase, the resonances for the lowest and highest frequency (2.33 GHz and 17.89 GHz for the glassy phase; 1.15 GHz and 19.88 GHz for the β-phase) are displayed. The solid black line represents the result of a global fit with multi-frequency dependent double-Gaussian derivatives that model both low and high-frequency data. All fits reveal the superposition of a broad and a narrow Gaussian. (c) Plots of the widths $\varDelta B$ of the two Gaussians for the two phases as a function of the applied on-resonance magnetic field $B$ based on the fit results obtained from the global fit procedure. The shaded regions represent 95% confidence intervals resulting from the parameter uncertainties that were determined using a bootstrap analysis. The circles and triangles represent the values of the continuous red and blue plots highlighted for the magnetic fields at which experimental spin resonance data were obtained.

# FIGURE 7

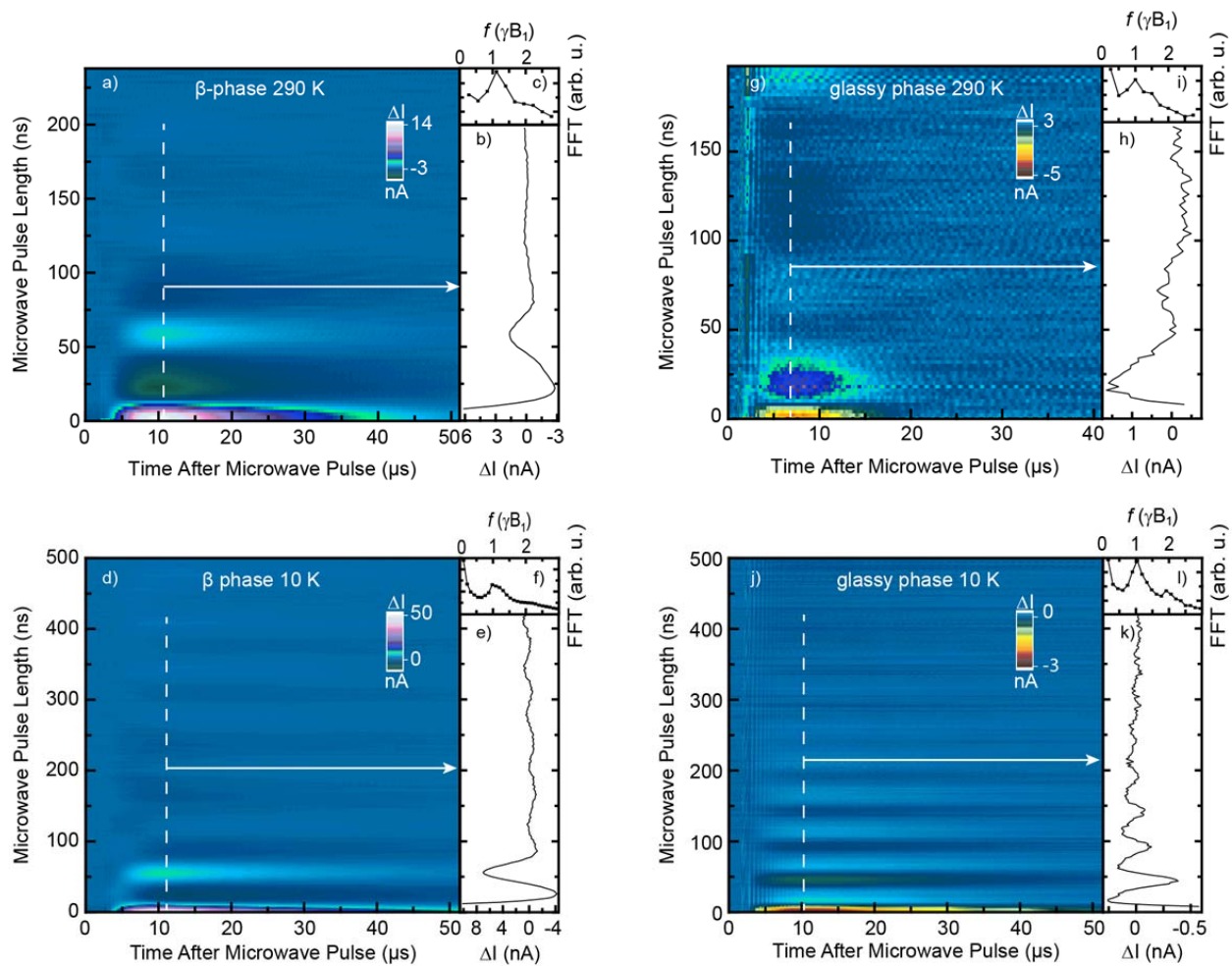

FIGURE 7: (a, d, g, j) Plots of the change in device current after a short, pulsed magnetic resonant excitation at $t=0$ as a function of time (horizontal axis) and pulse length (vertical axis) for both PFO phases at 293 K (top row) and 10 K (bottom row). The microwave pulse strength for all measurements was ~560 µT. The current transients are baseline corrected using a second-order polynomial to highlight the Rabi oscillations. (b, e, h, k) show corresponding slices along the dashed white line. (c, f, i, l) show the real component of the Fourier transforms that were calculated using data slices without background subtraction. The Fourier amplitude is plotted as a function of frequency in units of spin-½ Rabi frequency $\gamma B_1$, where $\gamma$ denotes the gyromagnetic ratio. Every frequency spectrum exhibits a peak in its signal intensity at the fundamental frequency of $\gamma B_1$ indicating the involvement of paramagnetic states with spin s=1/2. The 10 K measurements also show smaller peaks at $2\gamma B_1$, though less visible in the β-phase (f). This second harmonic is indicative of beating of the observed paramagnetic centers, the electron and hole spins.

# FIGURE 8

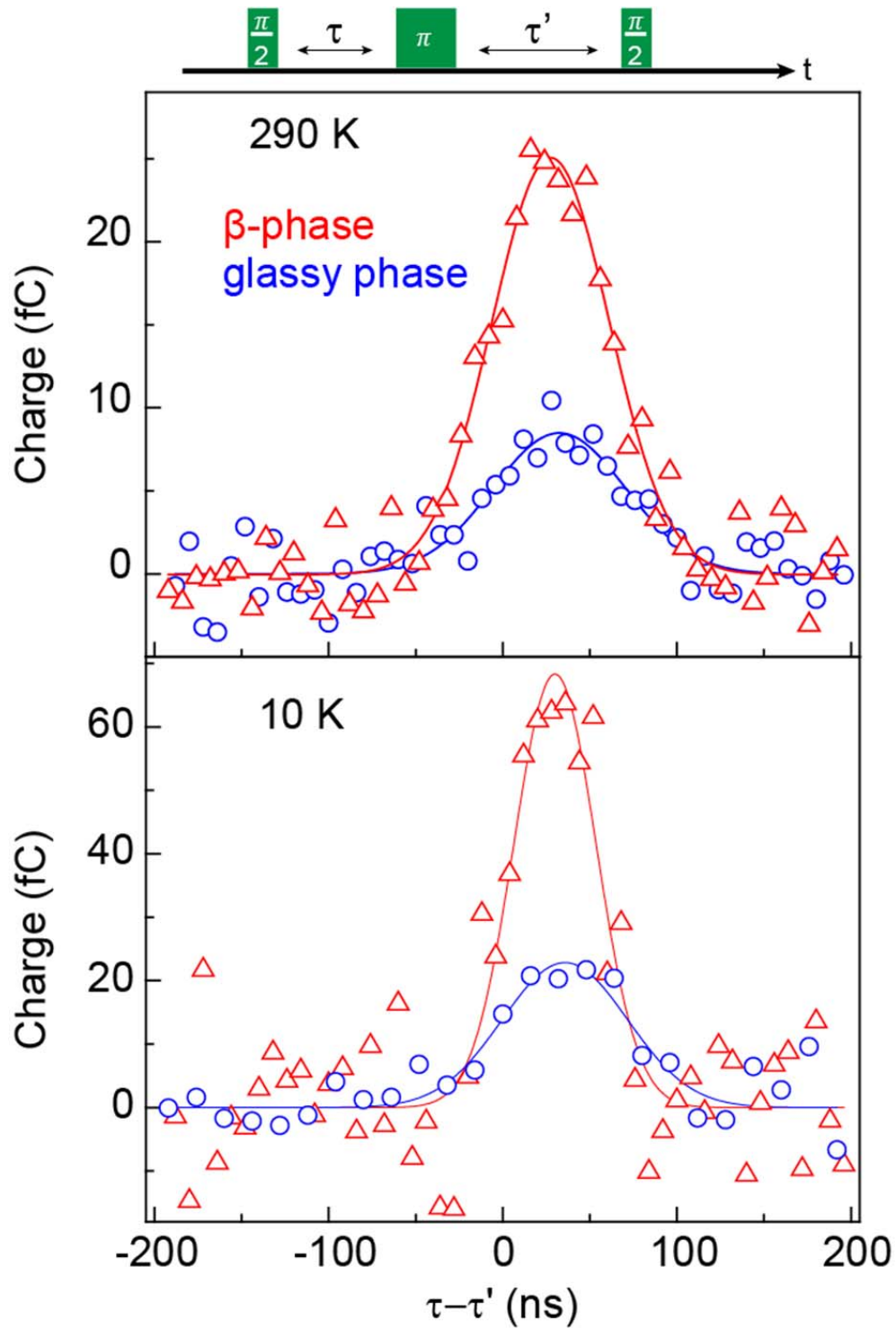

FIGURE 8: Plots of electrically detected Hahn echo experiments observed in the time-integrated current of devices with both PFO phases at 293 and at 10 K. The pulse sequence used for these experiments is sketched above the plots. Both plots show the integrated current (the charge) as a function of time difference $\tau - \tau'$ defined in the sketch of the pulse sequence. This difference was chosen such that the center of the electrically detected spin echoes occurs around $\tau - \tau' \approx 0$ for better comparison. The solid lines are fits with Gaussian curves and serve as a guide to the eye.

# FIGURE 9

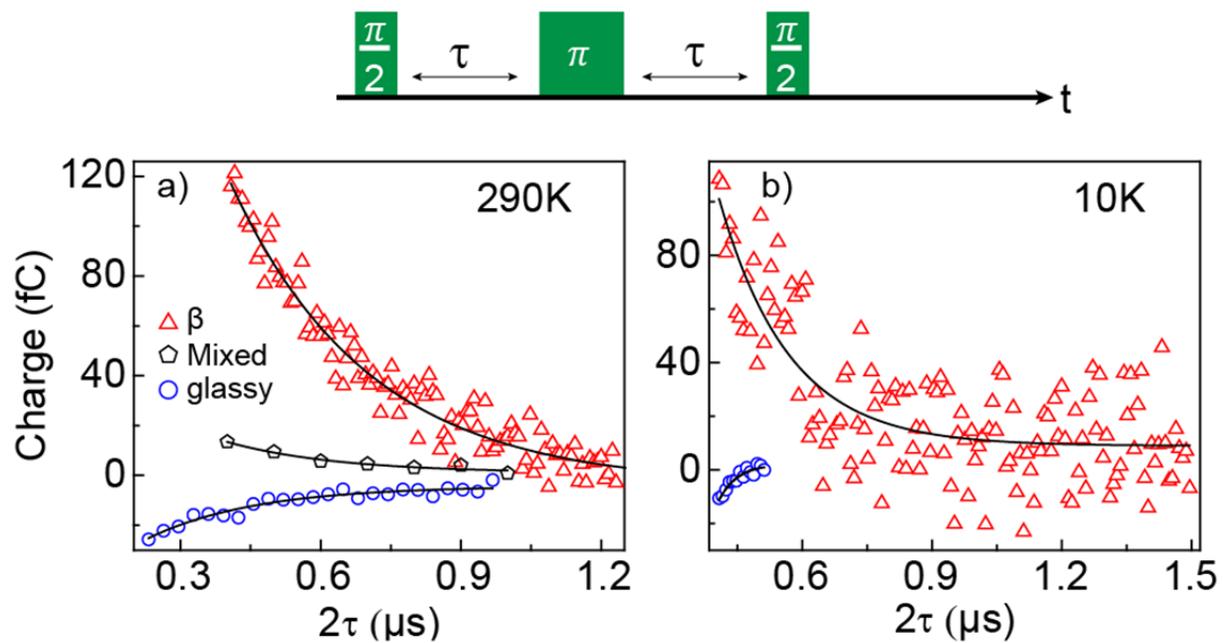

FIGURE 9: Plots of the decays of the Hahn echo envelopes measured at 293 K (a) and 10 K (b) as a function of $2\tau$ (defined in the pulse sequence diagram above). (a) Data were recorded from a mixed-phase device (black pentagons), a glassy-phase device (blue circles), and a β-phase device (red triangles). All measured data sets were fitted with single exponential decay functions in order to determine the coherence times $T_2$ of the charge carrier spin states.

# FIGURE 10

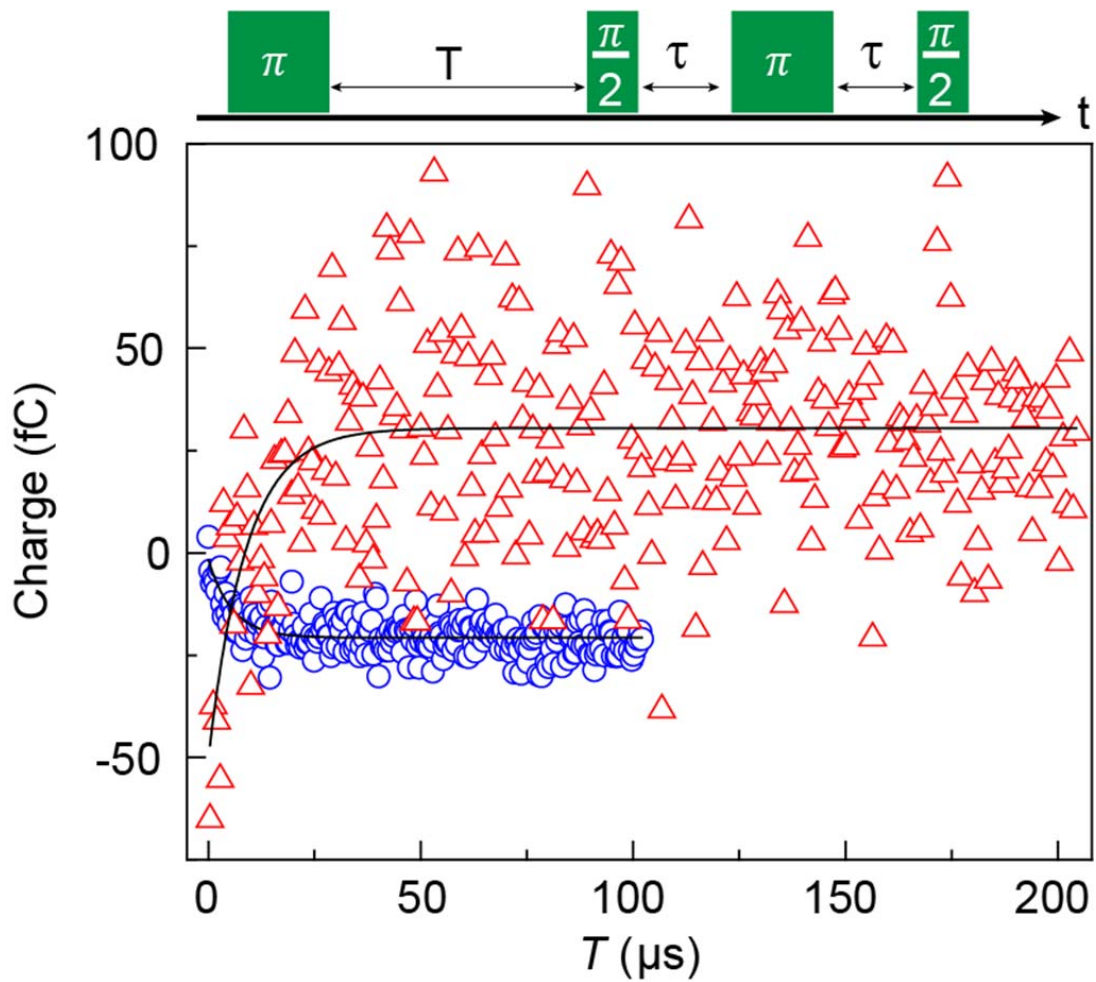

FIGURE 10: Plots of the results of inversion recovery experiments for glassy-phase (blue circles) and β-phase (red triangles) PFO OLEDs measured at 293 K. The time-integrated current (charge) is plotted as a function of mixing time (*T*) that follows the initial π pulse. An electrically detected Hahn echo sequence is used for readout. The EDMR pulse sequence used is shown above the plot. The β-phase OLED was forward biased so that the device current was 50 µA, with a circular active area of 500 µm diameter. The glassy device had a smaller active area with 200 µm diameter. It was operated at a current of $I_0 = 20$ µA.

# FIGURE 11

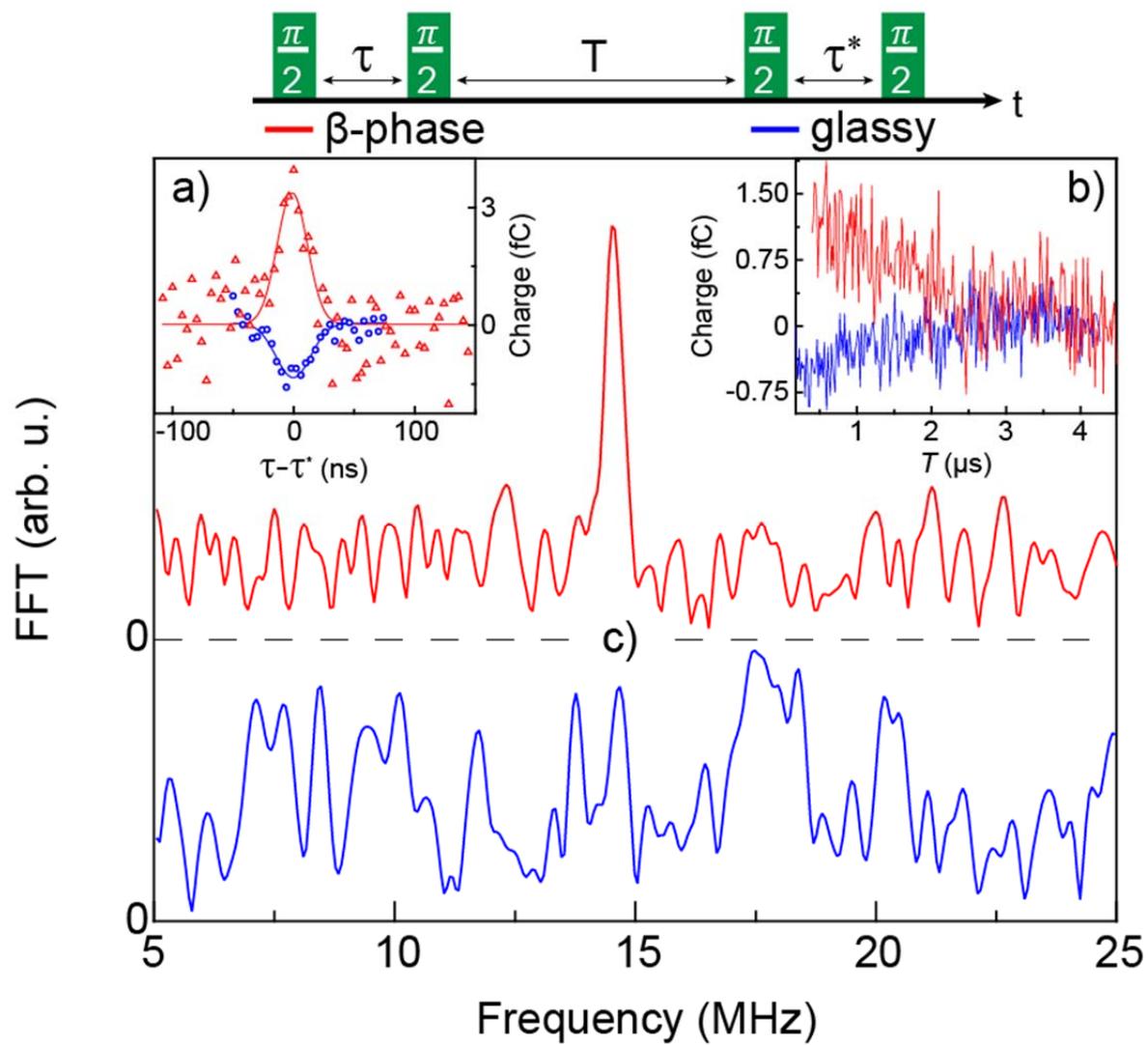

FIGURE 11: Plot of the stimulated and electrically detected echo measurements revealing the presence of an electron spin-echo envelope modulation (ESEEM) at room temperature. The definition of the pulse parameters $\tau$ and $T$ is shown in the sketch of the pulse sequence used. Blue and red data points represent results from glassy and β-phase devices, respectively. Panels a) and b) show the stimulated echoes and their decays for on-resonance measurements at a magnetic field of 344 mT. The stimulated echo is measured with a constant mixing time ($T = 180$ ns) and delay time $\tau$ while $\tau^*$ is varied (red and blue lines). For (b), $\tau = \tau^* = 96$ ns, with $T$ being varied. In (c) the Fourier transforms of the envelope modulations contained in the data in (b) are plotted, revealing a strong peak at 14.5 MHz for the β-phase device. This frequency corresponds to the proton Larmor frequency for the applied magnetic field. No significant modulation is seen in the glassy-phase device because of insufficient signal to noise.

# TABLE I

| β-phase | | Glassy phase | |
|---|---|---|---|
| Narrow line | Broad line | Narrow line | Broad line |
| $0.276 \leq B_{Hyp} \leq 0.280$ (mT) | $0.841 \leq B_{Hyp} \leq 0.851$ (mT) | $0.436 \leq B_{Hyp} \leq 0.446$ (mT) | $1.101 \leq B_{Hyp} \leq 1.120$ (mT) |
| $7.98 \leq \alpha \leq 8.17$ ($10^{-4}$) | $1.25 \leq \alpha \leq 1.31$ ($10^{-3}$) | $8.71 \leq \alpha \leq 9.43$ ($10^{-4}$) | $\alpha \leq 6.11 \times 10^{-4}$ |

TABLE I: Boundary values of the 95% confidence intervals for the double-Gaussian fit results of the multi-frequency c.w. EDMR data presented in Fig. 6. The ranges correspond to the shaded regions in Fig. 6(c). Line broadening arises due to both magnetic field-independent hyperfine coupling $B_{Hyp}$ and field-dependent broadening due to a distribution $\Delta g$ in Landé $g$-factors, denoted by the parameter $\alpha$. Hyperfine coupling is substantially stronger in the glassy phase than in the β-phase, even though the compounds are chemically identical. The broad line of the β-phase resonance shows significantly stronger broadening with magnetic field, suggesting that spin-orbit coupling may be stronger in the β-phase than in the glassy phase.

# TABLE II

|  | Decoherence time $T_2$ | | *Spin-lattice relaxation time $T_1$* |
|---|---|---|---|
|  | at 293 K | at 10 K | at 293 K |
| Glassy phase | 237 ± 37 ns | 590 ± 280 | 5.6 ± 0.9 (µs) |
| Mixed phase | 253 ± 82 ns | N/A | N/A |
| β-phase | 295 ± 10 ns | 252 ± 35 | 9.1 ± 2.5 (µs) |

TABLE II: Results of the bootstrap analysis of $T_1$ and $T_2$ times for the different phases of PFO. The errors stated represent one-sigma confidence intervals. The two-sigma values overlap for all phases, implying that there is no significant change in coherence time between the phases. The spin-lattice relaxation times for the β-phase appear to be slightly longer than for the glassy phase.